\PassOptionsToPackage{pdftex}{graphicx} 
\documentclass[10pt,conference]{IEEEtran}
\usepackage[pdftex]{graphicx}
\usepackage{latexsym}
\usepackage{multirow}
\usepackage{amssymb}
\usepackage{cite}
\usepackage{color}
\usepackage{balance}
\usepackage[hyphens]{url}
\usepackage{arydshln}
\usepackage{url}

\newcommand\sato[1]{}
\newcommand\para[1]{\noindent \textbf{{#1}}}

\begin{document}
%
\title{A Quantitative Study of Security Bug Fixes of GitHub Repositories}



\author{
\IEEEauthorblockN{
Daito Nakano\IEEEauthorrefmark{1},
Mingyang Yin\IEEEauthorrefmark{1},
Ryosuke Sato\IEEEauthorrefmark{1}, 
Abram Hindle\IEEEauthorrefmark{2},
Yasutaka Kamei\IEEEauthorrefmark{1} and
Naoyasu Ubayashi\IEEEauthorrefmark{1}
}
\IEEEauthorblockA{\IEEEauthorrefmark{1}Kyushu University\\Fukuoka, Japan\\\{nakano,yin,sato,kamei,ubayashi\}@posl.ait.kyushu-u.ac.jp}
\IEEEauthorblockA{\IEEEauthorrefmark{2}University of Alberta\\Edmonton, AB, Canada\\abram.hindle@ualberta.ca}
}



%


\maketitle

\begin{abstract}
  Software is prone to bugs and failures. Security bugs are those that
  expose or share privileged information and access in violation of the
  software's requirements. Given the seriousness of security bugs, there
  are centralized mechanisms for supporting and tracking these bugs
  across multiple products, one such mechanism is the Common
  Vulnerabilities and Exposures (CVE) ID description. When a bug gets a
  CVE, it is referenced by its CVE ID. Thus we explore thousands of
  Free/Libre Open Source Software (FLOSS) projects, on Github, to
  determine if developers reference or discuss CVEs in their code,
  commits, and issues. CVEs will often refer to 3rd party software
  dependencies of a project and thus the bug will not be in the actual
  product itself.  We study how many of these references are intentional
  CVE references, and how many are relevant bugs within the projects
  themselves. We investigate how the bugs that reference CVEs are fixed
  and how long it takes to fix these bugs. The results of our manual
  classification for 250 bug reports show that 88 (35\%), 32 (13\%), and
  130 (52\%) are classified into ``Version Update'', ``Fixing Code'',
  and ``Discussion''.
  To understand how long it takes to fix those bugs, we compare two
  periods, \emph{Reporting Period}, a period between the disclosure date
  of vulnerability information in CVE repositories and the creation date
  of the bug report in a project, and \emph{Fixing Period}, a period
  between the creation date of the bug report and the fixing date of the
  bug report. We find that 44\% of bug reports that are classified into
  ``Version Update'' or ``Fixing Code'' have longer Reporting Period than
  Fixing Period.  This suggests that those who submit CVEs should notify
  affected projects more directly.
 
\end{abstract}


%
\IEEEpeerreviewmaketitle

\section{Introduction}
\label{sec:intro}

In this study, we explore and characterize security bugs reported in
existing software systems.  Bug fixing is the main part of software
maintenance.  Since software is complicated and fault-prone, fixing
software bugs often requires the coordination of many actors, whose
actions are often unseen~\cite{Aranda09}. Also, once software is fixed,
the same problems, bugs, that were fixed in the past often
reoccur~\cite{Zimmermann12,ShihabEMSE2013}.  Furthermore, security
bugs~\cite{Zaman11} are often found to be more complicated than other kind of
bugs. This extra complication requires experienced developers to solve
these security bugs. Thus in this study, we focus on exploring how
developers address and fix security bugs, especially those that can
affect multiple products via their include dependencies.

Security bugs are often caused by bugs of the dependent programs.
When such a bug is addressed downstream, it is often via updating the library which had the security bug.
Many security fixes are dependency updates as the CVE is reported on a
dependency rather than the project itself.
This complicates root cause analysis, as Ma et al.~\cite{MaICSE17} have shown that it is difficult to diagnose the root cause of cross-project bugs.
Therefore, it is considered important to investigate how to fix bugs
that occurred in other development projects.

In order to distinguish security bugs from other kinds of bugs, we look for changes that refer \emph{Common Vulnerabilities and Exposures (CVE)}.
\emph{CVE} is a system for publicly known security vulnerabilities and exposures,
which provides a common identifier for security bug information.
CVE is used in both Free/Libre open source software (FLOSS) development and proprietary software.
Specific issues in CVE are named with CVE identifiers (e.g., CVE-2017-5753), which we call CVE IDs.
If a publicly known issue is addressed in CVE, vendors can communicate to their customers by using the information on CVE.
We conjecture that the bugs which refer CVE are security-related bugs.
We and others have found that the commits in Git repositories that fix the vulnerabilities related to CVE IDs often
mention the CVE IDs in their commit log messages.

When surveying bug fixes based on CVE description from Open Source Software (OSS) repositories, we found cases of serious security bugs that took
a long time to fix even after the project was informed of the bug.  For example, CVE-2014-6278 is a security bug of GNU bash, reported on
September 30, 2014.  On GitHub, 6 bug reports matched CVE-2014-6278 within one month of the CVE report (e.g., a bug report in
rapid7/metasploit-framework\footnote{\url{https://github.com/rapid7/metasploit-framework/issues/3931}}).  Even thought this issue is quite
serious, some projects that depends on \texttt{bash} did not update their dependencies immediately. For instance, by February 11, 2015, the Docker
project (e.g., a bug report in docker-library/official-images\footnote{\url{https://github.com/docker-library/official-images/issues/479}}) was
still distributing an official image that is vulnerable to the issue.

We investigate the artifacts extracted and we ask and answer 3 research questions to better understand the properties of bug fixes that refer CVE.
\begin{description}
 \item[RQ1] What kinds of security bug fixing patterns are there?
 \item[RQ2] How long will it take for security bugs to be fixed after being opened? 
 \item[RQ3] What kinds of products affect security bugs to ongoing projects?
\end{description}

These research questions are important to software development stakeholders and researchers.
Stakeholders can use these methods \sato{which methods?} from this survey to investigate how their projects depend on other projects that have handled security bug fixes.
Stakeholders can also understand what kind of cases they need to pay attention to the usage of specific types of products that are likely to have more security bugs.
Researchers can utilize the statistics from our study to estimate how many security bug fixes they can find by looking for CVE relevant commits.
Researchers can also leverage this work to find security-related bugs and insecure dependency related bugs.

The main contributions of this paper are:
\begin{itemize}
\item An empirically-grounded insight into the nature of security bugs in GitHub ecosystems that we derive from our quantitative and qualitative analysis (See the details of implication in Section~\ref{sec:conclusion}).
\item The definition of a classification scheme that describes the reasons for how GitHub projects deal with security bugs (RQ2).
\item All datasets and scripts used to conduct this study are available
      at \url{https://is.gd/8GeSTk}, which include the manual classification results of security bugs (RQ2 and RQ3). Researchers can
      replicate and extend our work using the datasets.
\end{itemize}

\begin{figure}[t]
  \centering
  \includegraphics[width=8cm]{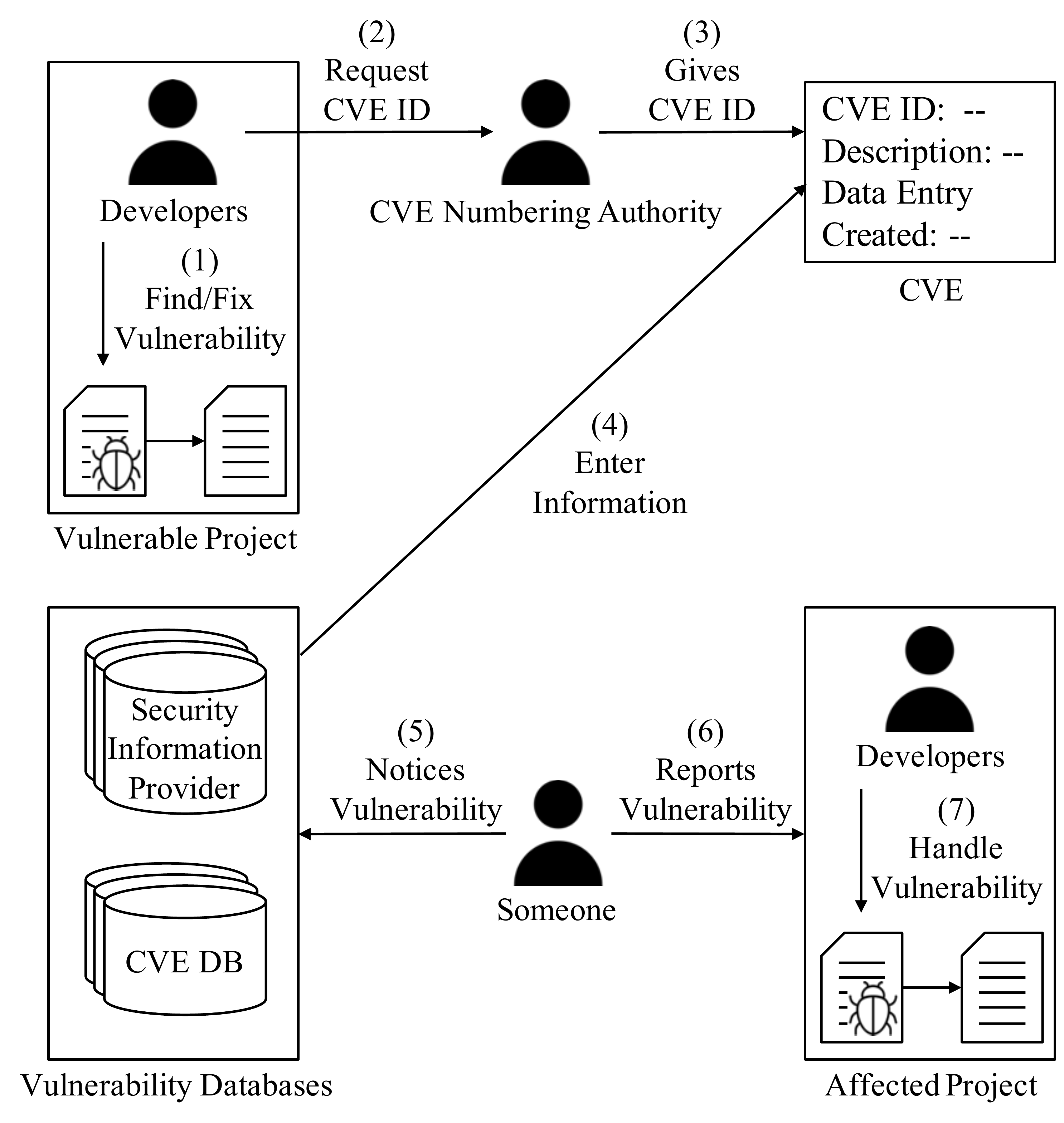}
  \caption{An example of how vulnerabilities information are handled}
  \label{fig:VulnerabilityFlow}
\end{figure}

The rest of the paper is organized as follows: Section~\ref{sec:back}
describes related work and Section~\ref{sec:dataset} describes datasets used in the paper.
Sections~\ref{sec:RQ1},\ref{sec:RQ2}, and \ref{sec:RQ3} pose 3 research questions and try to answer them
through manual and statistical analysis of the datasets. Then, Section \ref{sec:threats} discuss
possible threats to validity and \ref{sec:conclusion} concludes the paper.


\section{Background and Related Work}
\label{sec:back}
\subsection{Background}

\begin{figure}[t]
  \centering
  \fbox{
  \includegraphics[width=8cm]{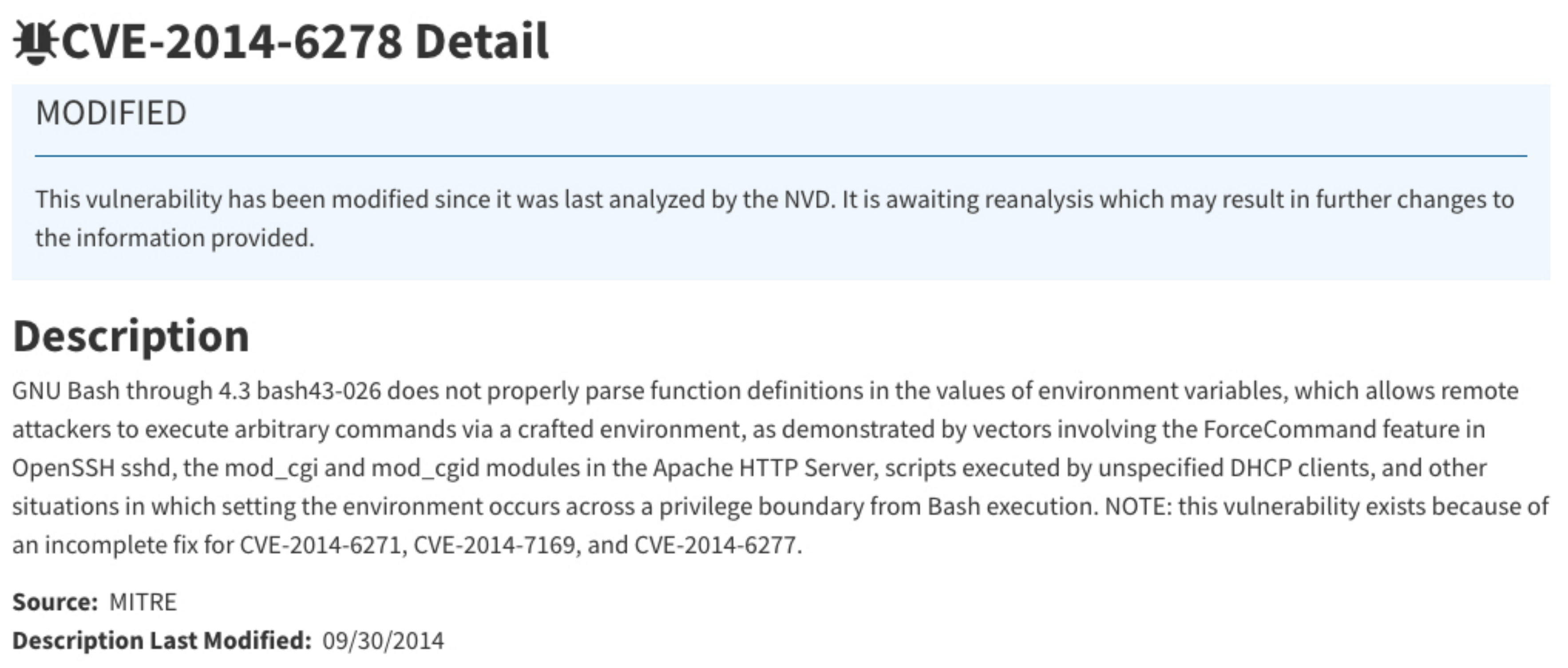}
  }
  \caption{A screenshot of CVE-2014-6278 informed from NVD}
  \label{fig:NVDscreenshot}
\end{figure}

\figurename~\ref{fig:VulnerabilityFlow} shows how vulnerability information is handled by using CVE.
When (1) software developers find a vulnerability in their project, they then fix the vulnerability.
After that, (2) they request a CVE Numbering Authority (CNA), an organization that create and announce CVE entries, to give a CVE ID for the vulnerability.
In \figurename~\ref{fig:VulnerabilityFlow}, a vulnerable project means a project that includes some vulnerability.
(3) CNA assign a CVE ID to the vulnerability, and (4) CVE databases (which store vulnerability information using CVE IDs as keys)
and security information providers (which store vulnerability information with or without CVE ID) enter the vulnerability information to their databases.
\figurename~\ref{fig:NVDscreenshot} shows a screenshot of a CVE informed from NVD, which is one of the CVE databases.
When (5) someone noticed the vulnerability information from these databases, and if they know another project affected by the vulnerability, then
(6) they report it to the project. In \figurename~\ref{fig:VulnerabilityFlow}, we call this project an affected project. 
(7) The developers of the affected project handle reported vulnerability (e.g., by using fixed versions).

Many studies analyze vulnerabilities by using some vulnerability databases and focus on how long vulnerable projects take time to fix vulnerabilities.
We focus on how affected projects handle vulnerabilities and how long it takes.

\subsection{Related Work}
Previous studies~\cite{Frei06, Frei08, Ransbotham10, Bilge12, ShahzadSL12, Gascon11, SarabiZXLD17, VasquezBE17} work on vulnerabilities extracted from vulnerability databases.
Some studies~\cite{Frei06, Frei08, Bilge12} focus on the time spent on fixing and exploiting vulnerabilities.
Other studies~\cite{Ransbotham10, ShahzadSL12} categorize vulnerabilities or projects to analyze whether the time spent on vulnerability fixing or exploiting depends on the categories.
There are also some studies~\cite{Gascon11, SarabiZXLD17, VasquezBE17} that use data sources other than vulnerability databases, such as Git repositories for applications on Android OS.

\para{Focus on the time of a vulnerability fixing or exploited.} Frei et al.~\cite{Frei06} quantified the
number of days from the day on which an exploitation is available to the day on which a patch is released.
They used CVE information collected from CVE databases (OSVDB~\cite{OSVDB} and NVD~\cite{NVD}) and
security information providers (CERT~\cite{CERT}, FrSirt~\cite{FrSirt}, ISS X-Force~\cite{ISSX-Force},
Secunia~\cite{Secunia} and SecurityFocus~\cite{SecurityFocus}) for quantifying the time. They analyzed the
numbers of days from 14,326 vulnerabilities. According to the analysis, exploit code is often created on the
day that the vulnerability is disclosed. Furthermore, they conducted an additional analysis by focusing on 2
vendors, Microsoft and Apple, between January 2002 and December 2007. They found that (1) Apple has more
unpatched vulnerabilities than Microsoft, and (2) Apple's rate of 0-day patches, patches released on the
vulnerability disclosure date, is between 0\% and 65\%, while Microsoft's is between 30\% and 90\%.
Bilge et al.~\cite{Bilge12} explored 0-day attacks on unreported vulnerabilities.
They identified 18 vulnerabilities that are exploited before reported, 11 of which have not been known as 0-day attacks.

\para{Categorize vulnerabilities or projects.}
Shahzad et al.~\cite{ShahzadSL12} extended Frei et al.'s work by adding a viewpoint of types of vendors and vulnerabilities. 
They classified vulnerabilities into some types and, for each type, they analyzed the number of days from the
day on which a vulnerability is disclose to the day on which a patch is released. The analysis showed that
most exploited types of vulnerabilities are denial of service, buffer overflow, and executable code.
Ransbotham~\cite{Ransbotham10} divided vulnerable projects into open-source software and closed-source
software. Regarding these two types of software, he then compared the opportunity of exploiting bugs before
they can be fixed.  He found that open-source software is attacked more frequently than closed-source
software.

\para{Use data source other than vulnerability databases.}
Gascon et al.~\cite{Gascon11} calculated delays from vulnerabilities information disclosure to updates to
Network Intrusion Detection Systems (NIDS).  They found that most of NIDS updates are released within 100
days after the disclosure of vulnerability information. They also found that NIDS updates faster if the severity of vulnerabilities is
higher.  Sarabi et al.~\cite{SarabiZXLD17} focused on software users. They investigated the number of days from the day on which
the vulnerable project publishes a patch to the day on which users apply the patch, by using WINE dataset~\cite{wine}.
Their data showed that users do not patch depending on the type of new release. \sato{???}
Linares-V\'asquez et al.~\cite{VasquezBE17} used CVE databases and the Git repositories for applications of Android OS.
They classified types of Android-related vulnerability, and found
that most vulnerabilities of Android OS are relevant to memory buffers, bug reports processing data, improper access control, and improper input validations.

\tablename~\ref{tb:reference_table} shows an overview of the related work.
In \tablename~\ref{tb:reference_table}, we use three viewpoints: (Time) Did they calculate the time of a
vulnerability fixing or exploiting? (Categorize) Did they categorize vulnerabilities or projects?
(Other data) Did they use data sources other than vulnerability databases?  For example,
Shahzad et al.~\cite{ShahzadSL12} calculated the number of days between vulnerabilities disclosure date and
patch available date, and they categorized vendors. Therefore, we check Time and Categorization
column. However, Shahzad et al. did not use data other than vulnerability databases. Thus, we do not check
Other data column.  Our studies calculated vulnerability fixing time of projects in GitHub, and categorize
the types of vulnerable projects, and use data of affected project mined from GitHub. Thus, we check every
column.

\tablename~\ref{tb:reference_table} shows that only Linares-V\'asquez et al.~\cite{VasquezBE17} and our studies have a checkmark in every column.
The difference between Linares-V\'asquez et al.~\cite{VasquezBE17} and our studies is the data sources used as vulnerability databases.
They used the Git data of Android OS. We, however, use the data of affected projects.
It means that the previous study used data between (1) and (4) in \figurename~\ref{fig:VulnerabilityFlow}, and we use data between (1) and (7). 
By using data between (1) and (7), we can analyze not only how long it takes vulnerable projects to fix a vulnerability but also how affected projects handle a vulnerability.

\begin{table}[t]
\centering
\caption{Overview of related work}
\begin{tabular}{l|ccc}
\hline
\multicolumn{1}{c|}{Paper} & Time & Categorize & Other data \\ \hline
Frei et al.~\cite{Frei06}&\checkmark&& \\
Frei et al.~\cite{Frei08}&\checkmark&& \\
Bilge et al.~\cite{Bilge12}&\checkmark&& \\
Shahzad et al.~\cite{ShahzadSL12}&\checkmark&\checkmark& \\
Ransbotham~\cite{Ransbotham10}&\checkmark&\checkmark& \\ 
Gascon et al.~\cite{Gascon11}&\checkmark&&\checkmark \\
Sarabi et al.~\cite{SarabiZXLD17}&&&\checkmark \\
Linares-V\'asquez et al.~\cite{VasquezBE17}&\checkmark&\checkmark&\checkmark \\ \hline
\textbf{Our study} &\checkmark&\checkmark&\checkmark \\ \hline
\end{tabular}
\label{tb:reference_table}
\end{table}

\section{Dataset}
\label{sec:dataset}

\begin{figure}[t]
  \centering
  \includegraphics[width=8cm]{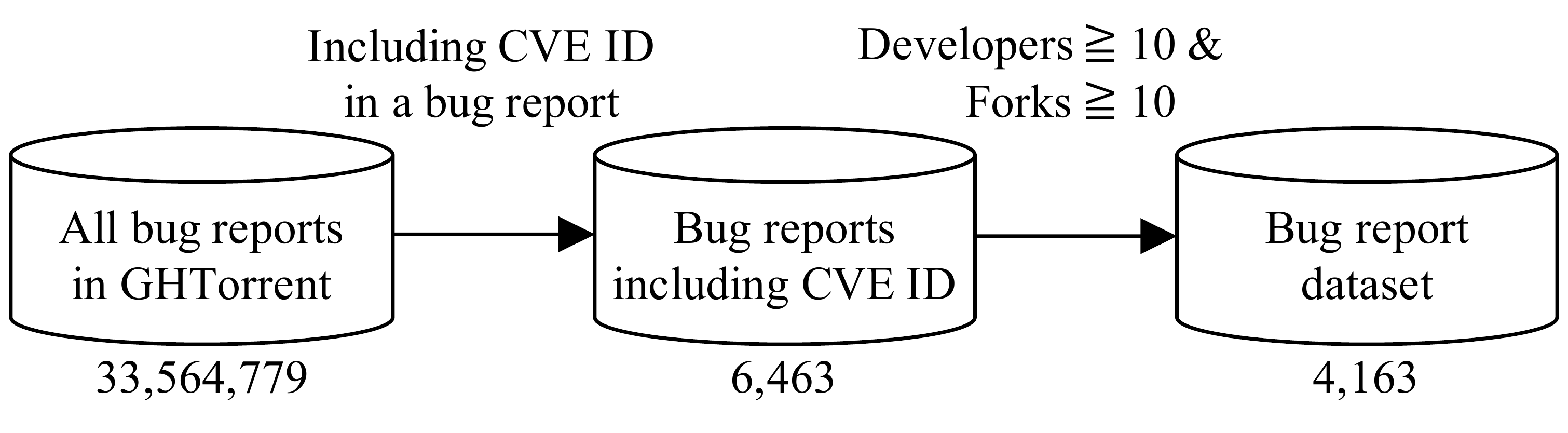}
  \caption{The number of bug reports of GitHub dataset.}
  \label{fig:github_dataset}
\end{figure}

To investigate bug fixes based on CVE description from OSS repositories,
we collected data from National Vulnerability Database (NVD) and GitHub.
In this section, we describe how we created our dataset from NVD and GitHub.

\subsection{CVE}
We collect CVE datasets (security bugs) from
NVD~\cite{NVD} as CVE DB in \figurename~\ref{fig:VulnerabilityFlow}. 
NVD archives meta-data of security issues, such as project names, release date, or vulnerability type, with CVE-ID. \sato{or -> and?}
As of July 15, 2017, we collected 87,066 security bugs from NVD.

\subsection{GitHub} \label{subsec:github}
GitHub is a popular open-source repository hosting service and contains
millions of projects.
GitHub project repositories include version control, bug trackers, wikis, and other services~\cite{EiriniMSR14}.

\begin{figure}[t]
  \centering
  \includegraphics[width=8cm]{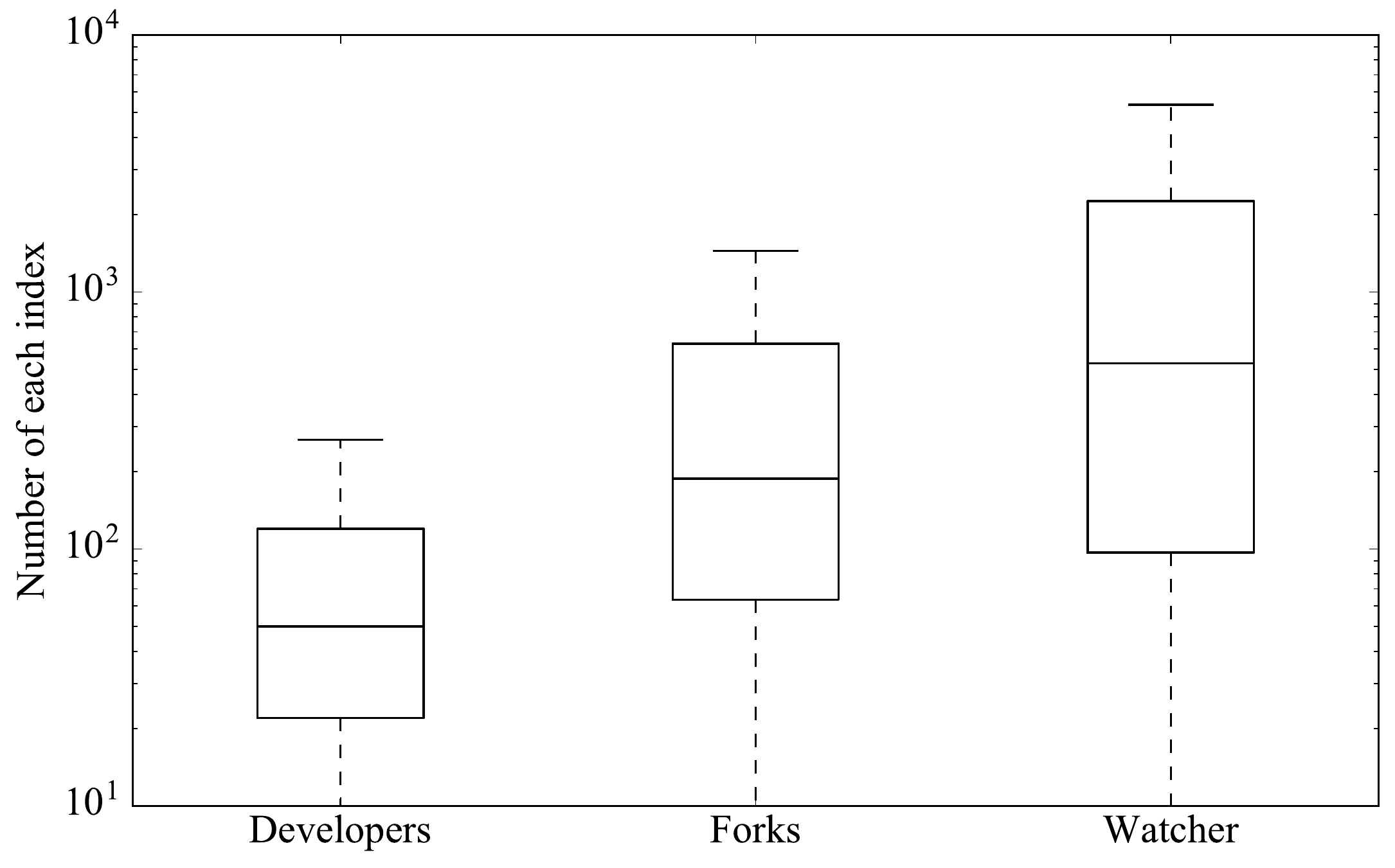}
  \caption{Distributions of project attributes studied in this work.}
  \label{fig:repodata}
\end{figure}

To find software projects that face security issues in GitHub, we used GHTorrent provided by Gousios et al.\cite{GeorgiosMSR13}.
GHTorrent is a database that aggregates and stores historical GitHub
project information and artifacts.
In this study, we focus on bug reports dealing with CVE information.
We extract only bug reports that include CVE-ID in titles, body texts, and discussions by using a regular
expression.\footnote{\textasciicircum(CVE\textbar cve)-[0-9]\{4\}-[0-9]\{4,5\}\$}

In this study, we select GitHub projects that develop software systems/applications, NOT homework,
documentation, and/or other non-software projects~\cite{MatragkasWKP14, SulirP16}.  We also exclude (1) forked
projects, which are working copy of the main repository and are created using the GitHub fork
feature\footnote{\url{https://help.github.com/articles/fork-a-repo/}}, and (2) projects with less than 10
developers or with less than 10 forks, since we want to target actively developed software systems.

\para{Data Overview.}  \figurename~\ref{fig:github_dataset} shows the number of entities within our dataset
extracted from GHTorrent.  The number of bug reports was 33,564,779, and 6,463 out of the bug reports contain CVE-ID in
their titles, descriptions, or discussions. After the filtering (1) forked projects as well as (2) projects
with less than 10 developers or with less than 10 forks, we finally obtain 4,163 bug reports that include
CVE-ID.

Overall, bug reports we analyze are collected from 1,304 projects. To better understand the overview of the
projects we analyze, \figurename~\ref{fig:repodata} shows the number of developers, the number of forks, and
the number of watchers (i.e., developers using the watch function to receive notifications of developing
activities) in the projects.  Of these repositories, there is a median of 50 developers, 188 forks, and 528
watchers. Many of these repositories are quite active.

\begin{figure}[t]
  \centering
  \includegraphics[width=8cm]{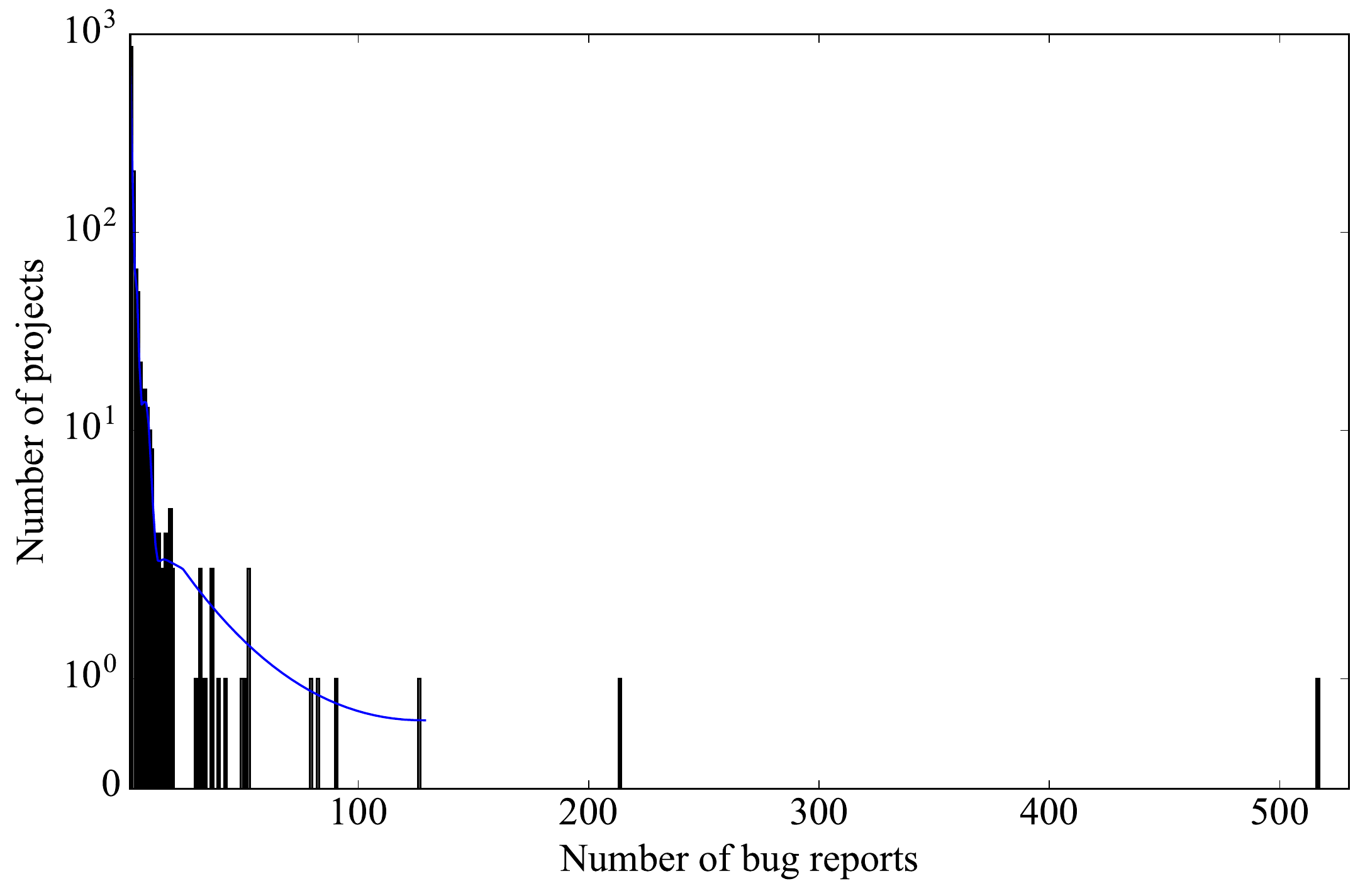}
  \caption{The number of bug reports across project including CVE ID.}
  \label{fig:NumofReports}
\end{figure}

\figurename~\ref{fig:NumofReports} shows how many bug reports including CVE IDs are reported in each project.
864 out of 1,304 projects have only one bug report including CVE IDs.  The largest number of bug reports in
one project is 516. This project is
metasploit/metasploit-framework\footnote{\url{https://github.com/rapid7/metasploit-framework}},
which provides vulnerability information and aids in penetration testing.

\begin{figure}[t]
  \centering
  \includegraphics[width=8cm]{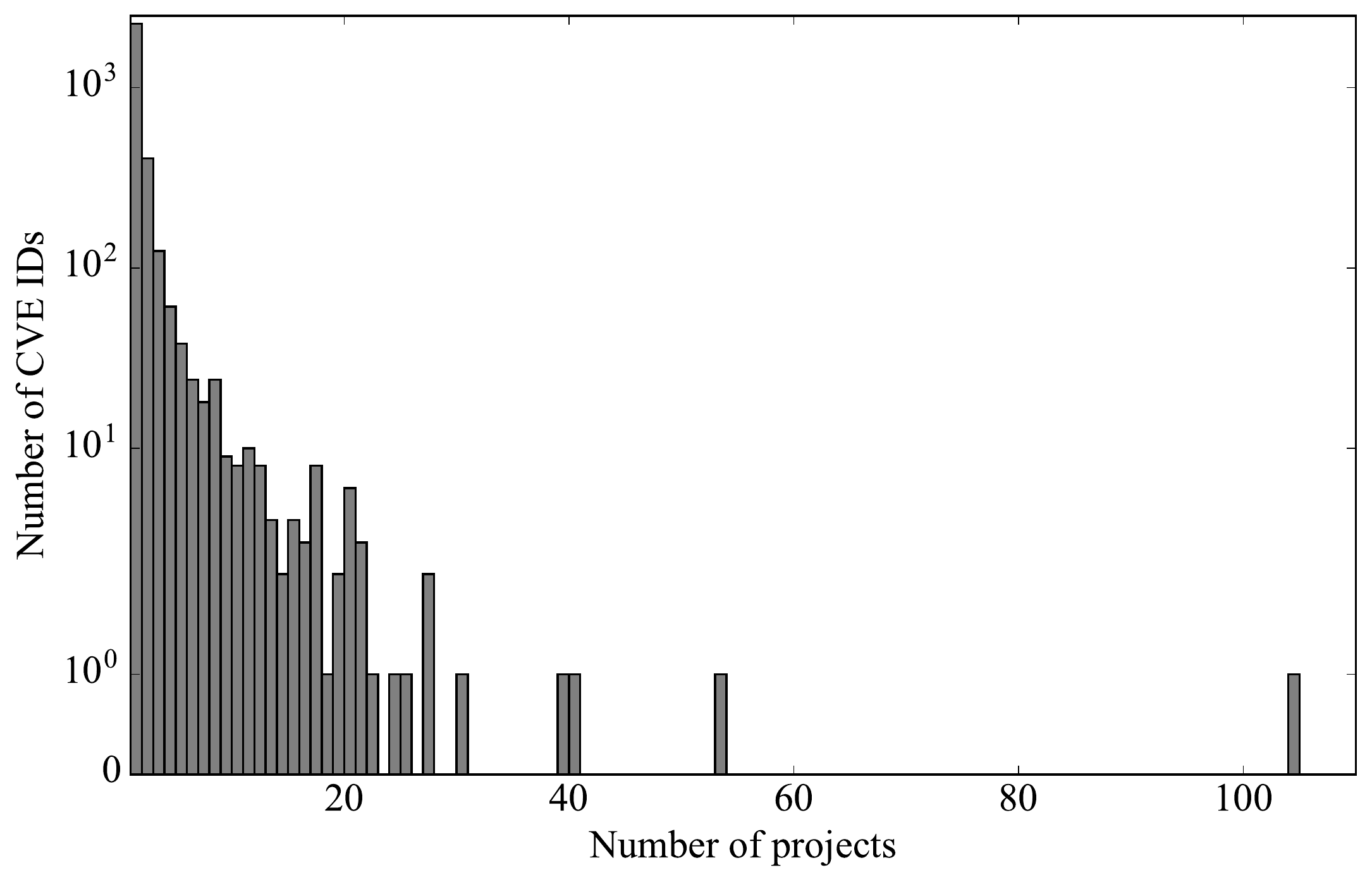}
  \caption{The number of projects that include same CVE ID.}
  \label{fig:NumofCVEID}
\end{figure}

The number of unique CVE IDs in 4,163 bug reports is 3,029.  \figurename~\ref{fig:NumofCVEID} shows how many
projects have the same CVE ID.  2,257 CVE IDs out of 3,029 are included in only one project.  One CVE ID
(CVE-2015-8103) is discussed across bug reports in 104 projects.  CVE-2015-8103 is the vulnerability that
occurs in the continuous integration tool named Jenkins.  This vulnerability allows remote attackers to
execute arbitrary code~\cite{CVE-2015-8103}.
This figure indicates that not all the CVE entries equivalently affect OSS development.

\section{(RQ1) What kinds of security bug fixing patterns are there?}
\label{sec:RQ1}
\subsection{Motivation}
\label{sec:1moti}

At least, 1,304 projects discuss security bugs. If there are bug fix patterns for security bug reports, such patterns
can be applied to fix future security bugs.
Therefore, we would like to manually
classify bug reports with CVE ID in order to discover what kinds of
effort are being spent and how security bugs are being addressed.

\subsection{Approach}
We randomly sample closed bug reports mentioned in Section~\ref{sec:dataset}. Then, we link the bug reports to related
CVE entry that is collected from NVD as discussed in Section~\ref{subsec:github}. We also obtain the commits that fix
the bug report if it includes a URL to the commits.

We manually classify these bug reports based on how security bugs were fixed/discussed. To perform classification,
we read the discussion in the bug reports, the security bug information obtained from NVD, the text, and source
code change of fixing commits. Then we classify bug reports manually.

In order to decide categories for classification, we conduct two iterations for manual classification. The
first iteration, the 1st author randomly samples and reads 70 bug reports from our dataset and creates
categories. Then, the 1st, 2nd, and 3rd authors discuss and decide three main categories (8 subcategories)
based on the 70 bug reports.

After that, we randomly sample 180 new bug reports resulting in 250 unique classified bug reports. The 1st
author classifies them. During this process, there are 11 bug reports that the 1st author is not confident
about and thus they require discussion. The first and second authors discussed the appropriate category for these
bug reports until they reached a consensus.

\begin{figure*}[t]
  \centering
  \includegraphics[width=18cm]{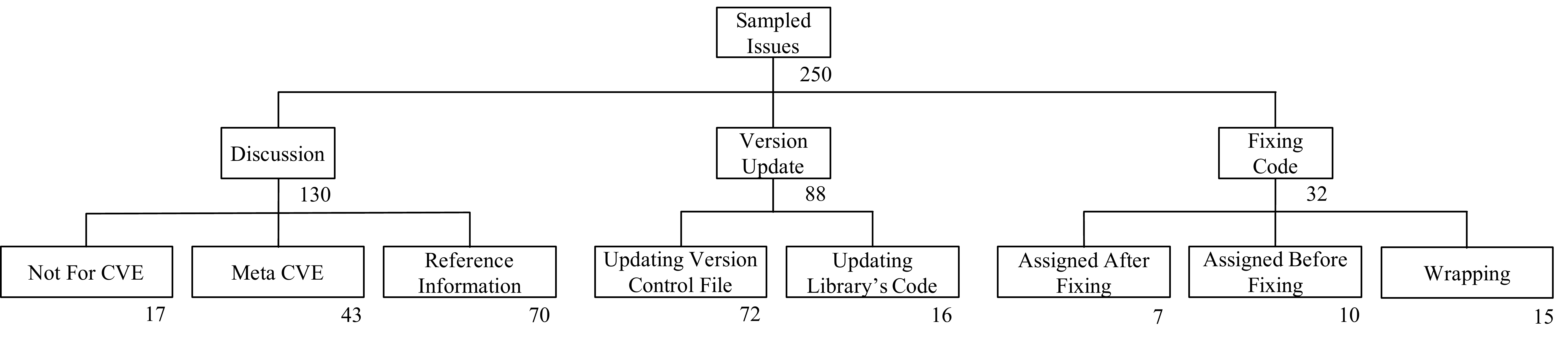}
  \caption{Security Bug Fixing Patterns}
  \label{fig:class}
\end{figure*}

\subsection{Result}

\para{Iteration 1.}
From 70 bug reports randomly sampled, we extract ``Fixing Code'', ``Version Update'', and ``Discussion'' as main categories.

We find that about half of the 70 bug reports actually fix security bugs. There are two major patterns (1) ``Version Update''
and (2) ``Fixing Code''. According to (1), a project notices that it faces security bugs caused by software libraries from other projects. In this case, the project fixes the security bugs by (1-1) modifying version numbers in version control files (e.g., pom.xml and Gemfile) or (1-2) updating/replacing the libraries from a buggy version to nonbuggy one. 
We call (1-1) and (1-2) as ``Updating Version Control File'' and ``Updating Library's Code''.

According to (2), we find that a project modifies its source code files because the security bugs are caused by (2-1) its own project or (2-2) other projects. In the case (2-1), developers modify the source code files in their project. We find that CVE-ID is given before/after
the bug report is closed.
We call each of them as ``Assigned Before Fixing'' and ``Assigned After Fixing''. As the case (2-2), we find that a project modifies its source code files to fix the security bugs that are caused by other projects, similar to (1) ``Version Update''. However, in the case (2-2), the project fixes its own source code files to avoid the security bug, NOT version control files nor updating libraries. We call the case 
``Wrapping''.


We find that half of the 70 bug reports do not fix security bugs. We classify these bug reports into ``Discussion''.
There are three subcategories: (1)
``Reference Information'' means that bug reports are closed without changing any source code files (e.g., a project opens a bug report to discuss one CVE-ID but the CVE actually does not affect the project),
(2) ``Meta CVE'' means that a project develops the tool (e.g., Metasploit~\cite{metasploit}) to scan vulnerabilities and
creates new bug reports for enhancement (e.g., scans for new CVE-IDs),
(3) ``False Positive'' means that the CVE-ID that is mentioned in the title, body text, and discussion is not related to the main topic of the bug report.

\para{Iteration 2.}
\figurename~\ref{fig:class} shows the results of our classification for 250 bug reports.

\noindent
\textit{\underline{Version Update.} }
Version updates are fixes that involve updating the versions of dependencies.
We find that 88 (35\%) out of 250 bug reports are classified into ``Version Update''. This means that 35\% of security bugs are caused by other projects and fixed by using the fixed version. To do so, 72 out of 88 security bugs are fixed by modifying version numbers in version control files.
For example, a developer of NixOS/nixpiks reported that a library used in NixOS/nixpiks was vulnerable.\footnote{\url{https://github.com/NixOS/nixpkgs/issues/6567}}
5 days after being reported, the developer modified a version control file called ``default.nix'' to fix the vulnerability.

16 out of 88 security bugs are fixed by updating/replacing libraries from a buggy version to a non buggy one. 
For example, a developer of nodejs/node reported that OpenSSL will release a new patch to fix security bugs (CVE-2015-1788).\footnote{\url{https://github.com/nodejs/node/issues/1921}}
After release, they fix the security bug by updating OpenSSL within one day of being released.

In common with two examples in ``Updating Version Control Files'' and ``Updating Software Libraries'', when comparing with other categories, developers are less likely to have discussions. When the developers notice that the software libraries that they use include security bugs, they just update the fixed version of the software including the security bugs.

\noindent
\textit{\underline{Fixing Code.} } These bug reports are fixed by modifying the projects' own source code to fix the bug.
We find that 32 (13\%) out of the 250 bug reports are classified into ``Fixing Code''. 
Among them, 17 bug reports are caused by security bugs in their own project.
For example, a user reported a security bug to developers in the-tcpdump-group/tcpdump.\footnote{\url{https://github.com/the-tcpdump-group/tcpdump/issues/446}} This security bug had not been assigned CVE ID at the reported time in the NVD database. The developer fixed the bug by modifying the project's source code files and added CVE-2015-3138 in 
the title of the bug report after CVE ID is assigned. 

Interestingly, the number of bug reports in the ``Fixing Code'' category is not too large. One reason that we guess is that developers discuss how to fix the security bugs in non-public communication channels such as a dedicated security mailing list. For example, a developer in theforeman/foreman made a pull request for fixing a security bug, and in this pull request, the developer described CVE ID for the bug\footnote{\url{https://github.com/theforeman/foreman/pull/373}}.
However, we cannot find reports or discussions about this bug before making the pull request.
It is because theforeman/foreman using private emails~\cite{foreman} for discussing security bugs, and developers discuss this security bug using private email before making this pull request.

We find that 15 bug reports are classified into the ``Wrapping'', which means that these bug reports are caused by other projects but fixed by changing their own projects' source code files.  
For instance, one bug report in mysociety/alaveteli\footnote{\url{https://github.com/mysociety/alaveteli/issues/2575}} is classified into ``Wrapping''.
The project uses Ruby on Rails and is affected by the Ruby on Rails's vulnerability, CVE-2015-3226.
A developer of mysociety/alaveteli made JSON initializer as a workaround to fix CVE-2015-3226.


\noindent
\textit{\underline{Discussion.} } These bug reports discuss issues around CVE-IDs but do not fix code or change dependencies.
We find that 130 (52\%) out of 250 bug reports are classified as ``Discussion''. In this category, bug reports contain CVE-IDs as references, but they are actually not fixed. Developers often use CVE-IDs for discussing whether the CVE-ID's security bugs are related to their projects.

In the ``Discussion'' category, there are 70 (54\%) bug reports as ``Reference Information''. Developers discuss whether or not the security bug should be fixed by pointing out its CVE-ID. 
For example, in the case of SeleniumHQ/selenium\footnote{\url{https://github.com/SeleniumHQ/selenium/issues/1375}}, a developer explains that one security bug (CVE-2015-4852) that affects Apache Commons Collections was disclosed. 
Because SeleniumHQ/selenium uses this library, the developer suggests that the library should be updated. 
A reply to this suggestion is that the library was updated before, and this bug report was closed. 
``Reference Information'' includes bug reports that contain discussion about a security bug and whether it is necessary to be fixed. Based on the discussion, developers typically find it unnecessary to fix.

We find that 43 (33\%) bug reports are related to ``Meta CVE''. ``Meta CVE'' includes bug reports that reported for vulnerability scanning tools. These projects deal CVE information not to fix but in order to exploit CVEs or find instances of CVEs in existing code. For example, in the case of wpscanteam/wpscan\footnote{\url{https://github.com/wpscanteam/wpscan/issues/330}}, a repository of a WordPress vulnerability scanner, a reporter requested to add a function detecting CVE-2013-4454 vulnerability.
This vulnerability (CVE-2013-4454) is in the WordPress plugin (portable-phpmyadmin) and allows direct access to a file without authority.
Based on the request, the function was actually implemented. 

17 (13\%) bug reports in the Discussion are ``False Positive'', i.e., a bug report includes CVE-IDs, but the main topic is not related to the CVE-IDs. Why this happens is because the CVE-IDs are included in the outputs of some tools. For example, issue \#197 of ouspg/trytls\footnote{\url{https://github.com/ouspg/trytls/issues/197}} shows the execution outputs of trytls. One line of 50 lines of the execution outputs includes ``PASS protect against Apple's TLS vulnerability CVE-2014-1266 [reject www.ssllabs.com:10443].'' The bug report was fixed but the CVE-ID (CVE-2014-1266) was not a main topic in the report. 

\smallskip
\begin{center}
\fbox{
\begin{tabular}{p{7.6cm}}
The results of our manual classification for 250 bug reports show that 
88 (35\%), 32 (13\%), and 130 (52\%) are classified into ``Version Update'', ``Fixing Code'', and ``Discussion''.
\end{tabular}
}
\end{center}
\smallskip


\section{(RQ2) How long will it take for security bugs to be fixed after being opened?}
\label{sec:RQ2}
\subsection{Motivation}
When a security bug occurs in the library that a development project uses, it may take time to notice that the security bug has occurred, but it does not take much time to be fixed. In such cases, we can reduce the fixing time of security bugs by notifying the occurrence of a bug in the library. Our example in RQ1 shows that the report of NixOS/nixpiks was fixed within 5 days, but however the vulnerability (CVE-2014-3618) described in the report was published by NVD about 1 year before the reported date.\footnotemark[4]
However, it is unknown whether this works. 
Hence, we investigate how long security bugs take time to be fixed after being opened.

\subsection{Approach}
We divide the whole period, from information disclosure of security bugs to the fixing, into two periods, because we would like to investigate the impact of reducing the period between the date of a security bug is disclosed and the date of the bug is reported.
The first period, the \emph{Reporting Period}, is a period between the date of vulnerability information disclosure in CVE repositories and the date of the bug report that is created on a project. 
The other period, the \emph{Fixing Period}, is a period between the date of the bug report that is created on the project and date of the bug that is fixed (i.e., the bug report is closed). The reporting period is calculated between the date that the CNA (3) Gives a CVE ID and the date that someone (6) Reports Vulnerability. The fixing period is calculated between the date of (6) and the date that a developer of the affected project (7) Handle Vulnerability in \figurename~\ref{fig:VulnerabilityFlow}.
We use the release date of CVE as the date of vulnerability information disclosure. 

The result of RQ1 shows that about half of the sampled bug reports actually do not fix security bugs, they only discuss the security bugs. 
To understand how long a security bug is fixed, we decided to use only the 120 bug reports which are categorized into ``Fixing Code'' or ``Version Update'', NOT use the 130 bug reports which are  categorized into ``Discussion''.

\begin{table}[t]
\centering
\caption{The number of days during reporting period and fixing period}
\scalebox{1}{
\begin{tabular}{c|c|rrrrr}
\hline
\multicolumn{1}{c|}{\multirow{2}{*}{Period}} & \multicolumn{1}{c|}{\multirow{2}{*}{Category}} & \multicolumn{5}{c}{Days} \\ \cline{3-7}
& & \multicolumn{1}{c}{Min.} & \multicolumn{1}{c}{.25}  & \multicolumn{1}{c}{Med.} & \multicolumn{1}{c}{.75} & \multicolumn{1}{c}{Max} \\ \hline \hline
Reporting& Version Update & -271 & -1 & 2 & 39 & 976 \\
Period& Fixing Code & -478 & -38 & -1 & 69.5 & 1823\\ \hline
Fixing& Version Update & 0 & 0 & 0 & 3 & 864\\
Period& Fixing Code & 0 & 0.5 & 1 & 14 & 610\\ \hline
\multicolumn{7}{l}{.25, .75 = interquartile range}
\end{tabular}
}
\label{tb:fixing_time}
\end{table}

\subsection{Result}
\tablename~\ref{tb:fixing_time} shows the number of days during the reporting period and the fixing period. 


\para{Reporting Period.}
Interestingly, we see negative values in the median column of Fixing Code in \tablename~\ref{tb:fixing_time}. It indicates that 50\% of bug reports are opened before a security bug is disclosed through NVD. For example, in the bug reported in the-tcpdump-group/tcpdump\footnote{\url{https://github.com/the-tcpdump-group/tcpdump/issues/446}{https://github.com/the-tcpdump-group/tcpdump/issues/446}}, an unknown security bug was reported and was fixed by developers. After fixing, developers required assigning CVE-ID for this bug, and CVE-2015-3138 was assigned 27 days after fixing. This suggests that  vulnerable projects, which the security bugs that occurred within its own development, are likely to report the bug reports after fixing the code.

We can see that the median time for Fixing Code is 1 day shorter than Version Update. 
We assume that it is because vulnerable projects are likely to have more security bugs ``Fixing Code'' than ``Version Update'' and be able to quickly report security bugs for Fixing Code.
However, we conduct exploit Mann-Whitney test with results and the result indicates that there is no statistically significant difference at $p = 0.27$. 

\para{Fixing Period.}
\tablename~\ref{tb:fixing_time} shows that Version Update is shorter than Fixing Code in terms of the median. We exploit Mann-Whitney U test with results indicated that there is statistically significant difference at $p < 0.01$. In Version Update row, the median of the number of days is 0. It means that more than half of bug reports are fixed within the reported day.

\para{Comparing Reporting Period with Fixing Period.}
We are interested in the bug reports that have reporting periods longer than fixing periods, because such bug reports may reduce more than half of the whole period of security bugs by notifying the existence of security bugs. 
When we focus on Version Update, there are 71\% of bug reports that the Reporting Period more than 0 days. There are 50\% of bug reports that the Reporting Period is longer than the Fixing Period. 
Similar to this trend, when we focus on Fixing Code, there are 42\% of bug reports that the Reporting Period is more than 0 days. There are 9\% of bug reports that the Reporting Period is longer than the Fixing Period. 

We would like to introduce an example that can shorten the latency period of security bugs.  
In the bug report of tricycle/predictionbook\footnote{\url{https://github.com/tricycle/predictionbook/issues/63}{https://github.com/tricycle/predictionbook/issues/63}}, developers fixed a security bug that occurred in rails by updating the version. 
This security bug was disclosed through NVD on 11/18/2014 and the developer made a pull request for fixing this bug on 2/17/2015. 
This pull request was accepted within the requested date. 
The Reporting Period of this bug is 93 days and the Fixing Period is 0 days. 
Like this case, if the Reporting Period is long and the Fixing Period is short, we can reduce the latent time of security bugs by notifying that bug information has been released.

\smallskip
\begin{center}
\fbox{
\begin{tabular}{p{7.6cm}}
The Reporting Period of 44\% of bug reports is longer than the Fixing Period. 
This suggests that they who submit CVEs should notify affected projects more directly.
\end{tabular}
}
\end{center}
\smallskip

\section{(RQ3) What kind of products affect security bugs to ongoing projects?}
\label{sec:RQ3}
\subsection{Motivation}
In this RQ, we classify the types of products that introduce security bugs discussed in bug reports. The findings of
this question have implications on the way how developers manage security bugs. For example, if projects face more
security bugs when they use one specific type of product, then the projects need to pay special attention to the usage
of the products. If it is not the case, the problem is less troubling.



\begin{table*}[t]
\centering
\caption{The percentage of domains each category}
\begin{tabular}{c|rrrrrr}
\hline
\multicolumn{1}{c|}{Category} & \multicolumn{1}{c}{Application} & \multicolumn{1}{c}{System} & \multicolumn{1}{c}{Web} & \multicolumn{1}{c}{Non-Web} & \multicolumn{1}{c}{Tools} & \multicolumn{1}{c}{Documentation} \\ \hline \hline
Version Update & 3 (3\%) & 45 (48\%) & 29 (31\%) & 10 (11\%) & 7 (7\%) & 0 (0\%) \\
Fixing Code & 5 (16\%) & 10 (36\%) & 10 (36\%) & 3 (10\%) & 3 (10\%) & 0 (0\%) \\ \hdashline
All & 8 (6\%) & 55 (44\%) & 39 (31\%) & 13 (10\%) & 10 (8\%) & 0 (0\%) \\ \hline
Borges et al.'s results~\cite{Borges16} & 269 (11\%) & 103 (4\%) & 837 (33\%) & 641 (26\%) & 270 (19\%) & 180 (7\%) \\ \hline
\end{tabular}
\label{tb:domain-category}
\end{table*}

\subsection{Approach}
We identify the types of vulnerable products that affected projects use. 
First, we create a list of CVE-IDs that are shown in the 120 bug reports that are classified into ``Version Update'' or ``Fixing Code''. We do not use 130 bug reports classified into ``Discussion'', because we want to analyze vulnerable projects that are actually fixed in affected projects. Then, we use the field of ``Affected Product'' in the NVD database to identify the name of products that have security bugs.  
We find that 269 unique CVE-IDs included in the 120 bug reports and 149 unique vulnerable products from the 269 CVE-IDs. Some projects have more than 1 CVE-IDs.

Borges et al.~\cite{Borges16} classified the top 2,500 public repositories in order of the number of stars in GitHub
into six domains. Following Borges et al., we use the following categories to classify the repositories:

\begin{itemize}
 \item Application software: systems that provide functionalities to end-users, such as browsers and text editors.
 \item System software: systems that provide services and infrastructure to other systems, such as operating systems, middleware, servers, and databases.
 \item Web libraries and frameworks, such as Django and Webrick.
 \item Non-web libraries and frameworks, such as OpenSSL.
 \item Software tools: systems that support software development tasks, such as IDEs, package managers, and compilers.
 \item Documentation: repositories that mainly provide documentation such as teaching materials and tutorials.
\end{itemize}

We classify 149 products into the six categories above.
We search an official web page using keywords (e.g., the product name) and read the contents of the official web page.
The first author is engaged in this classification task.
In this process, the official web pages could not be found for 11 products. We exclude the 11 products from this RQ.
The first author marked the seven products as low confidence for classification.
The seven products were discussed by the first and second authors until they reached a consensus decision.
As a result, we categorized 138 products.

\subsection{Result}
\tablename~\ref{tb:domain-category} shows the number of products within each domain. The category ``All'' shows the sum
of three categories. For comparison, the table also includes Borges et al.'s results~\cite{Borges16}. Interestingly, we
see a statistically significant difference between Borges et al.'s results and our results of ``All'' column
(Chi-squared test, $P < 0.001$).  While the percentage of system domain in our results shows 44\%, Borges et al.'s one
shows 4\%. Moreover, we see a different trend for non-web products. Our results indicate a smaller percentage
than Borges et al.'s results (10\% vs 26\%). This result implies that, when developers use system products for their
projects, they need to pay special attention to the usage of the products, for example, by checking the product's web page frequently to avoid
vulnerability bugs.


We also find that there is no documentation product. While the main contents of GitHub repositories sometimes provide
documentation such as teaching materials and tutorials, such products are less likely to affect other projects in terms
of vulnerable bugs.

\smallskip
\begin{center}
\fbox{
\begin{tabular}{p{7.6cm}}
While a previous study shows that the percentage of system software (4\%) is lowest in terms of popularity, our findings show that the percentage of system software (48\%) is highest in terms of vulnerability bugs. 
\end{tabular}
}
\end{center}
\smallskip

\section{Threat to Validity}
\label{sec:threats}
We now discuss the threat to the validity of our study.

\para{Construct validity.} considers the relationship between theory and observation, in case the measured variables do not measure the actual factors.
We looked for changes that refer CVE IDs. CVE ID is a common identifier for security bug information and it is often used in OSS and proprietary software.
However, we may miss security bugs that do not mention CVE IDs.
We need to confirm what percentage of bug reports mentions security bugs without CVE-IDs, which is left for future work.

\para{Internal validity.} refers to whether the experimental conditions make a difference or not, and whether there is
sufficient evidence to support the claim being made. To understand how long it takes to fix security bugs, we used the
release date of CVE as the date of vulnerability information disclosure from CVE DB.  It may happen that other channels
release vulnerability information earlier in public than CVE DB. To mitigate this threat, we used NVD, one of major CVE
DBs.

To conduct RQ2 and RQ3, we conducted manual classification to bug reports. Like any human activity, our manual classification results are biased and subjective. To mitigate this threat, we discussed classification results until they reach a consensus for bug reports that the 1st author is not confident.

\para{External validity.} considers the generalization of our findings.
We exclude projects that have less than 10 developers or less than 10 forked with other projects to remove immature projects~\cite{Yamashita15, EiriniMSR14}.
Although it is a common practice in previous studies, our results may not generalize to software projects with a small
number of contributors. In addition, while we target at only projects hosted in GitHub, other software ecosystems
such as SourceForge and BitBucket may have different trends.

\section{Conclusion}
\label{sec:conclusion}

In this paper, we investigate the security bugs described by CVE ID, based on three research questions (RQs). 

\emph{First}, we investigate what kinds of security bug fixing patterns there are. Most bug reports are addressed by updating the versions of software dependencies (``Version Update''), fixing the source code within the current project to deal with the vulnerability (``Fixing Code''), or a thorough discussion which results in no action taken to dependencies or source code due to irrelevance of the vulnerability.

\emph{Second}, we compared the Reporting Period and the Fixing Period. The Reporting Period is the period between the date of vulnerability information disclosure and the date of the bug report that is created on a project. The Fixing Period is the period between the date of the bug report that is created on the project and the date of the bug that is fixed.
As the result, we find that the Reporting Period of 44\% of ``Version Update'' or ``Fixing Code'' bug reports is longer than the Fixing Period---that is the time between a public CVE issuance and a bug report being created is larger than the bug report fix time. This suggests that those who submit CVEs should notify affected projects more directly.


\emph{Third}, we investigate what categories of products are affected by security bugs. We use the six categories that were defined by a previous study~\cite{Borges16}. We find that System software products, systems that provide services and infrastructure to other systems such as middleware, are typically affected by many CVE related issues (48\%). OSS developers should pay attention to especially the usage of System software products in their own projects.

\noindent {\bf Implications and Future Work.} 
The result of RQ1 shows 52\% of 250 bug reports are classified into ``Discussion'' that results in no action taken for the vulnerability. The result suggests that researchers who mine bug reports need filtering processes if they want to understand how to fix source code files for security bugs. Future works should investigate the approach that automatically detects whether or not a bug report actually results in action by using natural language processing, which is used in other tasks of software engineering studies~\cite{Maldonado_TSE2017}.

The results of RQ2 and RQ3 show that (1) time between a public CVE issuance and a bug report being created is larger than the bug report fix time and (2) System software products such as middleware and databases are more likely to be affected by CVE related issues. Although those who submit CVEs should notify affected projects more directly, those who use such products, especially System software products, should often check CVE DB.
One important future work is the impact of software bots~\cite{Lebeuf_IEEESoft2018} on CVE notification to OSS developers.


Future works should investigate whether similar results can be obtained from security bug fixes where CVE description does not exist. Additionally, we plan to better understand why some security bug reports take longer time than others by conducting interviews.

\section*{REPEATABILITY}
To enable repeatability of our work, and invite future research, we
provide all datasets and scripts that have been used to conduct this
study at \url{https://is.gd/8GeSTk}.

\bibliographystyle{IEEEtranS}
\bibliography{BugFix}

\balance

\end{document}